\documentclass[a4paper,12pt]{article}
\usepackage{t1enc}
\usepackage[latin1]{inputenc}
\usepackage[english]{babel}
\usepackage{graphics}
\begin{document}

\centerline {\bf Cosmoparticle Physics - the Challenge to the Millenium} 
\vskip0.2in
\centerline {\bf M.Yu. Khlopov$^{1}$}
\centerline {\it $^1$ Center for Cosmoparticle Physics "Cosmion",} 
\centerline {\it 125047, Moscow, Russia}
\vskip0.4in

\centerline{\bf Abstract}

Cosmoparticle physics offers the exciting challenge for the new Millenium 
to come to the true knowledge on the basic natural laws and on the way, they 
govern the 
creation and evolution of the Universe with all the forms of its present
energy content. The theory of everything, the true history of the Universe, 
based on it, the
new sources of energy release and the new means of energy transfer should 
come from the prospects of cosmoparticle physics. We may be near the 
first positive results in this direction. The methods of cosmoparticle 
physics are briefly reviewed.

\section{Introduction}

The modern cosmology implies A) inflation, B) baryosynthesis and C) dark matter 
and energy, what inevitably leads to the whole ABC, with which the NATURE writes 
the true WORDS in the BOOK (BIBLIO) on the History and Fate of the Universe. 
Cosmoparticle physics is aimed to find these Words, and to face on without fear 
the WORD, which was put into the Beginning of Creation. Treating the Universe 
as the physical process, the methods, developed by cosmoparticle physics, 
resolve the main problem of the modern cosmology: they probe the true picture of 
the cosmological evolution together with the whole set of physical laws, 
governing it. Based on the fundamental relationship between micro- and macro 
worlds, cosmoparticle physics offers the cross-disciplinary studies of the 
foundations of particle physics and cosmology in the proper combination of their 
indirect cosmological, astrophysical and physical signatures. 

Cosmoparticle physics originates from the well established relationship between 
microscopic and macroscopic descriptions in theoretical physics. Remind the 
links between statistical physics and thermodynamics, or between electrodynamics 
and theory of electron. Turning to the role of microscopic properties in 
macroscopic phenomena or to the investigation of microscopic processes by their 
macroscopic effects, we, in fact, indirectly use the ideas of cosmoparticle 
physics. In this sense all the methods of experimental study of elementary 
particles are based on these ideas, as well as these ideas underlie the 
theoretical description of astrophysical processes. To the end of the XX Century
the new 
aspects of such links were realised in cosmological necessity for the extension 
of the world of known elementary particles and their interactions and in the 
necessity for particle theory to use cosmological tests as the unique way
to probe its predictions.  

Being the last word of the fundamental physics of XX Century, cosmoparticle 
physics offers to the new millennium the set of questions. Their proper 
formulation is more than half of their proper answers. There are serious grounds 
to expect in the very near future definite answers on some of these questions. 
It specifies the modern step of development of cosmoparticle physics: on the basis 
of its principles and in its framework it not only 
sheds new light and offers nontrivial solutions for old astronomical problems 
but also comes to positive predictions of new phenomena, accessible for 
their search.

\section{The "ecological" aspects of the modern cosmology}

Ecology studies the mutual relationship between the population and its 
environment. In this sense the mutual relationship between the fundamental 
particle content, particle interactions and the structure and evolution of the 
Universe, studied by cosmoparticle physics, is ecological. 

Lets specify the links between fundamental particle properties and their 
cosmological effects in some more details. The role of particle content in the 
Einstein equations is reduced to its contribution into energy-momentum tensor. 
So, the set of relativistic species, dominating in the Universe, realises the 
relativistic equation of state $p= \epsilon/3$ and the relativistic stage of 
expansion. The difference between relativistic bosons and fermions or various 
bosonic (or fermionic) species is accounted by the statistic weight of 
respective degree of freedom. The very treatment of different species of 
particles as equivalent degrees of freedom physically assumes strict symmetry 
between them.

Such strict symmetry is not realised in Nature. There is no exact supersymmetry 
(symmetry between bosons and fermions). There is no exact symmetry between 
various quarks and leptons. The symmetry breaking results in generation of 
particle masses. The particle mass pattern reflects the hierarchy of symmetry 
breaking.

Noether's theorem relates the exact symmetry to conservation of respective charge. The lightest particle, bearing the strictly conserved charge, is absolutely stable. So, electron is absolutely stable because of the conservation of electric charge. The mass of electron $m_{e}$ is related to the scale of the 
electroweak symmetry breaking, $\Lambda_{EW}$: the standard model gives the 
electron mass as $m_{e}= g_{e} \Lambda_{EW}$, where $g_{e}$ is the Higgs field 
Yukawa coupling. In the same manner the stability of proton is conditioned by 
the conservation of baryon charge and its mass reflects the chiral symmetry 
breaking being determined by the scale of QCD confinement, $\Lambda_{QCD}$. The 
stability of ordinary matter is thus protected by the conservation of electric 
and baryon charges, and its properties reflect the fundamental physical scales 
of electroweak and strong interactions.

Extensions of the standard model imply new symmetries and new particle states. 
The respective symmetry breaking induces new fundamental physical scales in 
particle theory. If the symmetry is strict, its existence implies new conserved 
charge. The lightest particle, bearing this charge, is stable. The set of new 
fundamental particles, corresponding to the new symmetry, is then reflected in the existence of new stable 
particles, which should be present in the Universe and taken into account in the 
total energy-momentum tensor. 

Most of the known particles are unstable. For a particle with the mass $m$ the 
particle physics time scale is $t \sim \frac{1}{m}$, so in particle world we 
refer to particles with lifetime $\tau \gg \frac{1}{m}$ as to metastable. To be 
of cosmological significance metastable particle should survive after the 
temperature of the Universe $T$ fell down below $T \sim m$, what means that the 
particle lifetime should exceed $t \sim \frac{m_{Pl}}{m} \cdot \frac{1}{m}$. 
Such a long lifetime should find reason in the existence of an (approximate) 
symmetry. From this viewpoint, cosmology is sensitive to the most fundamental 
properties of microworld, to the conservation laws reflecting strict or nearly 
strict symmetries of particle theory.

However, the mechanism of particle symmetry breaking can also have the 
cosmological impact. Heating of condensed matter leads to restoration of its 
symmetry and to formation of topological defects in the course of phase 
transitions, corresponding to symmetry 
breaking. One can directly observe formation of such 
defects in liquid crystals or in superfluids. In the same manner the mechanism 
of spontaneous symmetry breaking should reflect in restoration of symmetry and,
depending on the symmetry breaking pattern, to formation of topological defects 
in phase transitions in very early Universe. The defects can represent the new 
form of stable particles (as it is in the case of magnetic monopoles), or the 
form of extended structures, such as cosmic strings or cosmic walls. In the 
latter case primordial strong non-homogeneity appears, triggering gravitational 
instability.

In the old Big bang scenario the cosmological expansion and its initial 
conditions was given {\it a priori}. The whole set of fundamental particles, 
most of which are unstable, was taken into account in the conditions of 
thermodynamic equilibrium, and the number of particle species entered the 
relationship between the temperature and cosmological time. The properties of stable 
particles cause much more influence on the physics of expansion. The baryon mass 
at initially given baryon to photon ratio determines the change of equation of 
state in the transition from radiation to matter dominance. The chemical 
composition, resulting from primordial nucleosynthesis is defined by the rate of 
beta-reactions, determining the neutron to proton frozen out ratio, by nuclear 
reaction rates and Coulomb interaction (defining the Coulomb barrier in 
reactions with electrically charged nuclei). The period of recombination of 
hydrogen is determined by electron mass and its Coulomb interaction and scales 
of gravitational instability is defined by the dissipation scale, being in turn 
determined by photon-electron Compton scattering. In the framework of standard 
model of particle interactions the above microphysical 
parameters reflect the QCD and electroweak scales. In the early Universe, when 
the temperature was of the order of these scales the QCD and electroweak phase 
transitions should have taken place.  
 
In the modern cosmology the expansion of the Universe and its initial conditions 
is related to the process of inflation. The global properties of the Universe as 
well as the origin of its large scale structure are the result of this process. 
The matter content of the modern Universe is also originated from the physical 
processes: the baryon density is the result of baryosynthesis and the 
nonbaryonic dark matter represents the relic species of physics of the hidden 
sector of particle theory. Physics, underlying inflation, baryosynthesis and 
dark matter, is referred to the extensions of the standard model, and the 
variety of such extensions make the whole picture in general ambiguous. 
However, in the framework of each particular physical 
realisation of inflationary model with baryosynthesis and dark matter the 
corresponding model dependent cosmological scenario can be specified in all the 
details. In such scenario the main stages of cosmological evolution, the 
structure and the physical content of the Universe reflect the structure of the 
underlying physical model. The latter should include with necessity the standard 
model, describing the properties of baryonic matter, and its extensions, 
responsible for inflation, baryosynthesis and dark matter. In no case the 
cosmological impact of such extensions is reduced to reproduction of these three 
phenomena only. The nontrivial path of cosmological evolution, specific for each 
particular realisation of inflational model with baryosynthesis and nonbaryonic 
dark matter, always contains some additional model dependent cosmologically 
viable predictions, which can be confronted with astrophysical data. The part of 
cosmoparticle physics, called cosmoarcheology, offers the set of methods and 
tools probing such predictions.   

\section{Cosmoarcheology of new physics in the modern cosmology}

Cosmoarcheology considers the results of observational cosmology as the sample 
of the experimental data on the possible existence and features of hypothetical 
phenomena predicted by particle theory. To undertake the {\it Gedanken 
Experiment} with these phenomena some theoretical framework for treatment of 
their origin and evolution in the Universe should be assumed. As it was pointed 
out \cite{Cosmoarcheology} the choice of such framework is a nontrivial problem 
in the modern cosmology. 

Indeed, in the old Big bang scenario any new phenomenon, predicted by particle 
theory was considered in the course of the thermal history of the Universe, 
starting from Planck times. The problem is that the bedrock of the modern 
cosmology, namely, inflation, baryosynthesis and dark matter, is also based on 
experimentally unproven part of particle theory, so that the test for possible 
effects of new physics is accomplished by the necessity to choose the physical 
basis for such test. There are two possible solutions for this problem: a) a 
crude model independent comparison of the predicted effect with the 
observational data and b) the model dependent treatment of considered effect, 
provided that the model, predicting it, contains physical mechanism of 
inflation, baryosynthesis and dark matter.

The basis for the approach (a) is that whatever happened in the early Universe 
its results should not contradict the observed properties of the modern 
Universe. The set of observational data and, especially, the light element 
abundance and thermal spectrum of microwave background radiation put severe 
constraint on the deviation from thermal evolution after 1 s of expansion, what 
strengthens the model independent conjectures of approach (a). 

One can specify the new phenomena by their net contribution into the 
cosmological density and by forms of their possible influence on parameters of 
matter and radiation. In the first aspect we can consider strong and weak 
phenomena. Strong phenomena can put dominant contribution into the density of 
the Universe, thus defining the dynamics of expansion in that period, whereas 
the contribution of weak phenomena into the total density is always subdominant. 
The phenomena are time dependent, being characterised by their time-scale,
so that permanent (stable) and temporary (unstable) phenomena can take place.
They can have homogeneous and inhomogeneous distribution in space. The amplitude 
of density fluctuations $\delta \equiv \frac{\delta \rho}{\rho}$ measures the 
level of inhomogeneity. The case $\delta \ge 1$ within the considered component 
corresponds to its strong inhomogeneity. Strong inhomogeneity is compatible with 
the smallness of total density fluctuations, if the contribution of inhomogeneous component 
into the total density is small. 

The phenomena can influence the properties of matter and radiation either indirectly, 
say, changing of the cosmological equation of state, or via direct interaction with 
matter and radiation. In the first case only strong phenomena are relevant, in the 
second case even weak phenomena are accessible to observational data. The detailed 
analysis of sensitivity of cosmological data to various phenomena of new physics 
are presented in \cite{book}.

The basis for the approach (b) is provided by a particle model, in which 
inflation, baryosynthesis and nonbaryonic dark matter is reproduced. 
Any realisation of such physically complete basis for models of the modern 
cosmology contains with necessity additional model dependent predictions, 
accessible to cosmoarcheological means. Here the scenario should contain all 
the details, specific to the considered model, and the confrontation with the 
observational data should be undertaken in its framework. In this approach 
complete cosmoparticle physics models may be realised, where all the 
parameters of particle model can be fixed from the set of astrophysical, 
cosmological and physical constraints. Even the details, related to 
cosmologically irrelevant predictions, such as the parameters of unstable 
particles, can find the cosmologically important meaning in these models. 
So, in the model of horizontal unification, the top quark or B-meson physics 
fixes the parameters, describing the dark matter, forming the large scale 
structure of the Universe.

\section{Cosmophenomenology of new physics}

To study the imprints of new physics in astrophysical data cosmoarcheology 
implies the forms and means in which new physics leaves such imprints. So, the 
important tool of cosmoarcheology in linking the cosmological predictions of 
particle theory to observational data is the {\it Cosmophenomenology} of new 
physics. It studies the possible hypothetical forms of new physics, which may 
appear as cosmological consequences of particle theory, and their properties, 
which can result in observable effects. 

The simplest primordial form of new physics is the gas of new stable massive 
particles, originated from early Universe. For particles with the mass $m$, at 
high temperature $T>m$ the equilibrium condition, $n \cdot \sigma v \cdot t > 1$ 
is valid, if their annihilation cross section $\sigma > \frac{1}{m m_{Pl}}$ is 
sufficiently large to establish the equilibrium. At $T<m$ such particles go out 
of equilibrium and their relative concentration freezes out. More weakly 
interacting species decouple from plasma and radiation at $T>m$, when $n \cdot 
\sigma v \cdot t \sim 1$, i.e. at $T_{dec} \sim (\sigma m_{Pl})^{-1}$. The 
maximal temperature, which is reached in inflationary Universe, is the reheating 
temperature, $T_{r}$, after inflation. So, the very weakly interacting particles
with the annihilation cross section $\sigma < \frac{1}{T_{r} m_{Pl}}$, as well as very heavy particles with the mass $m \gg T_{r}$ can not be 
in thermal equilibrium, and the detailed mechanism of their production should be considered to 
calculate their primordial abundance.

Decaying particles with the lifetime $\tau$, exceeding the age of the Universe, 
$t_{U}$, $\tau > t_{U}$, can be treated as stable. By definition, primordial 
stable particles survive to the present time and should be present in the modern 
Universe. The net effect of their existence is given by their contribution into 
the total cosmological density. They can dominate in the total density being the dominant form of cosmological dark matter, or they can represent its subdominant fraction. In the latter case more detailed analysis of their distribution in 
space, of their condensation in galaxies, of their capture by stars, Sun and Earth, as well as of the effects of their interaction with matter and of their annihilation provides more sensitive probes for their existence. In particular, direct experimental search for cosmic fluxes of weakly interacting massive particles (WIMPs) is possible. WIMP annihilation in galactic halo contributes into the fluxes of cosmic rays, and their annihilation in Sun and Earth is the source of neutrinos accessible to underground neutrino observatories. New particles with electric charge and/or strong interaction can form anomalous atoms and contain in the ordinary matter as anomalous isotopes.

Primordial unstable particles with the lifetime, less than the age of the Universe, $\tau < t_{U}$, can not survive to the present time. But, if their lifetime is sufficiently large to satisfy the condition $\tau \gg \frac{m_{Pl}}{m} \cdot \frac{1}{m}$, their existence in early Universe can lead to direct or indirect traces. Cosmological flux of decay products contributing into the cosmic and gamma ray backgrounds represents the direct trace of unstable particles. If the decay products do not survive to the present time their interaction with matter and radiation can cause indirect trace in the light element abundance or in the fluctuations of thermal radiation. If the particle lifetime is much less than $1$s the multi-step indirect traces are possible, provided that particles dominate in the Universe before their decay. On the dust-like stage of their dominance black hole formation takes place, and the spectrum of such primordial black holes traces the particle properties (ma!
!
ss, frozen concentration, lifetime). The particle decay in the end of dust like stage influences the baryon asymmetry of the Universe. So cosmophenomenoLOGICAL chains link the predicted properties of even unstable new particles to the effects accessible in astronomical observations. 

The parameters of new stable and metastable particles are determined by the pattern of particle symmetry breaking. This pattern is reflected in the succession of phase transitions in the early Universe. The phase transitions of the first order proceed through the bubble nucleation, which can result in black hole formation. The phase transitions of the second order can lead to formation of topological defects, such as walls, string or monopoles. The observational data put severe constraints on magnetic monopole and cosmic wall production, as well as on the parameters of cosmic strings. The succession of phase transitions can change the structure of cosmological defects. The more complicated forms, such as walls-surrounded-by-strings can appear. Such structures can be unstable, but their existence can lead the trace in the nonhomogeneous distribution of dark matter and in large scale correlations in the nonhomogeneous dark matter structures, such as {\it archioles}. The large sc!
!
ale correlations in topological defects and their imprints in primordial inhomogeneities is the indirect effect of inflation. Inflation provides the equal conditions of phase transition, taking place in causally disconnected regions. Phase transitions, taking place directly at inflational stage, can lead to strong inhomogeneity at any scale,
Which can lead to formation of primordial black holes of a whatever large mass. In the combination with successive phase transitions, taking place after reheating, phase transitions at inflational stage can result in such structures as closed walls of any size, collapsing into black holes after their size equals the horizon. So the primordial strong inhomogeneities is the new important phenomenon of cosmological models, based on particle models with hierarchy of symmetry breaking.

\section{Experimental probes for new physics}

The new physics follows from the necessity to extend the Standard model. The white spots in the representations of symmetry groups, considered in the extensions of the Standard model, correspond to new unknown particles. The extension of the symmetry of gauge group puts into consideration new gauge fields, mediating new interactions. Global symmetry breaking results in the existence of Goldstone boson fields. 

For a long time the necessity to extend the Standard model had purely 
theoretical reasons. Esthetically, because full unification is not
achieved in the Standard model; practically, because it contains some 
internal inconsistencies. It does not seem complete for cosmology. 
One has to go beyond the Standard model to explain inflation, baryosynthesis 
and nonbaryonic dark matter. Recently there has appeared a set of experimental 
evidences for the existence of neutrino oscillations and WIMPs, and for the 
effects of new particles in the precise measurements of muon magnetic 
momentum. Whatever convincing and reliable these evidences are, they 
indicate that may be we have already crossed the border in the experimental 
searches for new physics. 

In particle physics direct experimental probes for the predictions of particle 
theory are the most attractive. The predictions of new charged particles, 
such as supersymmetric particles or quarks and leptons of new generation, 
are accessible to experimental search at accelerators of new generation, 
if their masses are in 100GeV-1TeV range. However, the predictions related 
to higher energy scale need non-accelerator or indirect means for their test. 

The search for rare processes, such as proton decay, neutrino oscillations, 
neutrinoless beta decay, precise measurements of parameters of known 
particles, experimental searches for dark matter represent the widely 
known forms of such means.

Cosmoparticle physics offers the nontrivial extensions of indirect and 
non-accelerator searches for new physics and its possible properties. In 
experimental cosmoarcheology the data is to be obtained, necessary to 
link the cosmophenomenology
of new physics with astrophysical observations (See \cite{Cosmoarcheology}).
In experimental cosmoparticle physics the parameters, fixed from the consitency
of cosmological models and observations, define the level, at which the new
types of particle processes should be searched for (see \cite{expcpp}). 

Note that there is a way, in which new physics may be elusive for the 
existing methods of experimental particle physics: the higher is the energy
of colliding particles the higher is the resolution in small spatial scales,
but if some abnormal forms of particles exist, being extended in space, 
their existence can escape detection. For example, if some fictious 
spatially isomeric
form of proton existed, having the form of a rod with the molecular size length, 
it could hardly been detected by the means of deep inelastic collisions.
For the moment, there is no theoretical reasoning for such exotic forms of
known particles, but the principal possibility for such solutions can not be
ignored.  

\section{Cosmoparticle physics of theories of everything}

The theories of everything should provide the complete physical basis
for cosmology. The problem is that the string theory \cite{Green} 
is now in the form
of "theoretical theory", for which the experimental probes are widely doubted
to exist. The development of cosmoparticle physics can remove these doubts.
In its framework there are two directions to approach the test of 
theories of everything. 

One of them is related with the search for the experimentally
accessible effects of heterotic string phenomenology. The mechanism of 
compactification and symmetry breaking leads to the prediction of 
homotopically stable objects \cite{Kogan1} and shadow matter \cite{Kogan2},
accessible to cosmoarcheological means of cosmoparticle physics.
The condition to reproduce the Standard model naturally leads in the
heterotic string phenomenology to the prediction 
of fourth generation of quarks and leptons 
\cite{Shibaev} with a stable massive 4th neutrino \cite{Fargion99},
what can be the subject of complete experimental test in the near future.
Moreover, there are evidences from EGRET galactic 
gamma-background measurements
and underground WIMP searches, favoring the hypothesis of 4th neutrino 
with the mass about 50 GeV, and it was recently shown 
that capture and annihilation of such neutrinos and their antineutrinos
inside the Earth, should lead to the flux of underground monochromatic
neutrinos of known types, which can be traced in the analysis of the
already existing data of underground neutrino detectors \cite{Belotsky}.

It is interesting, that heterotic string phenomenology  predicts 
even in its simplest realisation both supersymmetric particles and the
4th family of quarks and leptons. Provided that both R-parity and the
new gauge charge, ascribed to the 4th generation, are strictly conserved,
the same model predicts simultaneously two types of WIMP candidates:
neutralinos and massive stable 4th neutrinos. So in the framework
of this phenomenology the multicomponent analysis of WIMP effects
is favorable. 

In the above approach some particular phenomenological features of 
simplest variants of string theory are studied. 
The other direction is to elaborate
the extensive phenomenology of theories of everything by adding to the symmetry
of the Standard model the (broken) symmetries, which have serious reasons to 
exist. The existence of (broken) symmetry between quark-lepton families,
the necessity in the solution of strong CP-violation problem with the use of
broken Peccei-Quinn symmetry, as well as the practical necessity 
in supersymmetry to eliminate the quadratic divergence of Higgs boson mass
in electroweak theory is the example of appealing additions to the symmetry
of the Standard model. The horizontal unification and its cosmology represent
the first step on this way, illustrating the might of cosmoparticle physics in
the elaboration of the proper phenomenology for theories 
of everything \cite{Sakharov}.

\section{Conclusion}

We can conclude that our ideas on the Universe experience the dramatic change, 
comparable with the one, caused by the Copernicus idea 
that Earth moves around Sun and by the Friedman's 
idea on the non-stationary Universe. From the very beginning to the modern 
stage, the evolution of Universe is governed by the forms of matter, different 
from those we are built of and observe around us. From the very beginning 
to the present time, the evolution of the Universe was governed by 
physical laws, which we still don't know. Observational cosmology offers 
strong evidences favouring the existence of processes, determined by 
new physics. Obsevations favor the dominance of new forms of matter in the total energy 
of the modern Universe.

Cosmoparticle physics, studying the physical, astrophysical and cosmological 
impact of new laws of Nature, explores the new forms of matter 
and their physical properties, what opens the 
way to use the corresponding new sources of energy and  
new means of energy transfer. It offers the great challenge for 
the new Millennium. 

It's regrettful, that A.D.Sakharov isn't now with us on this way, 
but he was with us in its beginning \cite{ADS}, and it's the aim of the present 
conference to present the new step in the development of Sakharov's
legacy in the field of cosmoparticle physics \cite{MKH}.

\section{Acknowledgement}

The work was performed in the framework of the project "Cosmoparticle physics"
and was partially supported by the Cosmion-ETHZ and AMS-Epcos collaborations
and by  a support grant for the Khalatnikov Scientific School.
The author expresses his gratitude to IHES for its kind hospitality.


\begin{thebibliography}{99}

\bibitem{Cosmoarcheology}  {\small M.Yu. Khlopov,
Cosmoarcheology. Direct and indirect astrophysical effects of 
    hypothetical particles and fields, In: Cosmion-94, Eds. M.Yu.Khlopov et al.
Editions frontieres, 1996. PP. 67-76}

\bibitem{book}  {\small M.Yu. Khlopov,
Cosmoparticle physics, World Scientific, 1999.}

\bibitem{expcpp}  {\small K.M.Belotsky, M.Yu. Khlopov, 
A.S.Sakharov, A.L.Sudarikov, A.A.Shklyaev,
Experimental cosmoparticle physics: experimental probes for 
    dark matter physics at particle accelerators, 
    Gravitation and Cosmology (1998), V. 4, Supplement.  PP. 70-78}

\bibitem{Green}  {\small M. Green, J. Schwarz, E. Witten, Superstring
theory, Cambridge University Press, 1989.}

\bibitem{Kogan1}   {\small Ia.I. Kogan, M.Yu. Khlopov,  Homotopically stable 
particles in superstring theory. 
    Yadernaya Fizika (1987) V. 46, PP. 314-316}

\bibitem{Kogan2}   {\small Ia.I. Kogan, M.Yu. Khlopov,  Cosmological 
consequences of superstring models. 
    Yadernaya Fizika (1986) V. 44, PP. 1344-1347}
    
\bibitem{Shibaev}   {\small M.Yu.Khlopov, K.I.Shibaev, This volume}

\bibitem{Fargion99}  {\small D. Fargion, Yu.A. Golubkov, M.Yu. Khlopov, R.V.
Konoplich, R. Mignani, Possible effects of the existence of 4th generation
neutrino, JETP Letters (1999) V.69 (N6), P.434 ; astro/ph-9903086}

\bibitem{Belotsky}   {\small K.M.Belotsky, M.Yu.Khlopov, This volume}

\bibitem{Sakharov}    {\small A.S.Sakharov, M.Yu.Khlopov, 
Horizontal unification 
as phenomenology of theory of "everything". 
    Yadernaya Fizika (1994) V. 57, PP. 690-697}
    
\bibitem{ADS}    {\small A.D.Sakharov, Cosmoparticle physics 
the cross-disciplinary science,
Vestnik AN SSSR (1989), V.4, PP. 39-40}

\bibitem{MKH}    {\small M.Yu.Khlopov, Fundamental cross-disciplinary 
studies of microworld and Universe. 
    Vestnik of Russian Academy of  Sciences (2001) V.71, PP. 1133-1137}

\end{thebibliography}
\end{document}